\documentclass[a4paper,11pt]{article}
\pdfoutput=1 

\usepackage{jcappub} 

\usepackage[T1]{fontenc} 

\title{\boldmath Wave function of the universe in the presence of trans-Planckian censorship}

\author[a]{Vikramaditya Mondal}

\affiliation[a]{School of Physical Sciences, Indian Association for the Cultivation of Science,\\Kolkata - 700032, India}

\emailAdd{vikramaditya.academics@gmail.com}

\abstract{The wave function for a closed de Sitter universe has been computed, demanding consistency with the recently proposed Trans-Planckian Censorship Conjecture (TCC). We extend the Einstein-Hilbert action to contain a complex-valued term which provides an exponentially decaying weight for the geometries violating TCC in the Lorentzian path integral sum while working in the minisuperspace approach to quantum cosmology. This \textit{postulated} modification suppresses the probability of evolution of the universe into configurations that violate TCC. We show that due to the presence of this suppression factor, the Hubble rate of the universe at the end of the inflation gets subdued and assumes a value less than what is expected classically. Moreover, the consequences of this quantum gravity motivated correction in the primordial power spectrum are discussed as well.}

\begin{document}
\maketitle
\flushbottom

\section{Introduction}

A surprising revelation in the post-nineteenth-century physics was the realization that the laws of quantum physics are more fundamental, of which the classical laws are crude approximations. Given this fundamental characteristic of quantum physics, it begs the question of whether quantum laws, instead of classical ones, should govern the evolution of the universe. Even though the present universe might not reveal any quantum behavior, the quantum nature is expected to be critical for its evolution in the early times.\par
Considering the entire universe as a quantum mechanical system is one of the boldest proposals in modern theoretical cosmology. Since the universe is considered a quantum system, it must be associated with a wave function. The most concrete attempts to compute such a wave function are Hartle-Hawking's ``no-boundary proposal'' \cite{hartle1983wave} and Vilenkin's ``tunneling universe proposal'' \cite{vilenkin1982creation}. These proposals consider the scenario of the `birth' of a de Sitter universe from apparently `nothing.' However, recent investigations \cite{obied2018sitter,bedroya2020trans} into string theory, a well-developed candidate for a theory of quantum gravitation, indicate that effective field theories leading to long-lived de Sitter evolution of the universe belong to the Swampland---a set of low energy effective physics models that are inconsistent with the theory of quantum gravity \cite{vafa2005string}. The proposed semiclassical wave functions remain oblivious to such a constraint on de Sitter spaces coming from quantum gravity. We seek an appropriate modification to the wave function of the universe such that the non-viability of long-lived de Sitter evolution is encoded into it and study further consequences.\par
The universe's wave function can be computed either from the path integral approach or by solving the Wheeler-DeWitt equation. For our purpose, the path integral approach seems to be most appropriate. Historically, the Euclidean path integral approach was used to derive the wave function (see, for example, \cite{hartle1983wave,halliwell1989steepest}). However, the Euclidean approach is \textit{not} devoid of mathematical and physical issues \cite{gibbons1977einstein,gibbons1978path}. On the other hand, given the recent development of evaluating oscillatory lapse function integrals using Picard-Lefschetz theory, a new approach dubbed Lorentzian Quantum Cosmology (LoQC) is being explored \cite{feldbrugge2017lorentzian}. Picard-Lefschetz theory removes the ambiguity in choosing the steepest descent contours and allows for systematic evaluation of either the Green's function or the wave function depending on the range for the lapse function integration. Due to the Lorentzian path integral's recognized rigor over the Euclidean approach, we should be using LoQC as the basis for our analysis here (for recent applications of LoQC to bouncing cosmological models, see \cite{PhysRevD.103.106008,rajeev2022bouncing}).\par
Moreover, recent attempts to re-derive Hartle-Hawking's and Vilenkin's wave function within the fold of LoQC have created a controversy \cite{feldbrugge2017lorentzian,feldbrugge2017no,dorronsoro2017real,dorronsoro2018damped,vilenkin2018tunneling,feldbrugge2018no,feldbrugge2018inconsistencies,vilenkin2019tunneling}. Later, it has been proposed that to avoid running into the problem of uncontrolled perturbations around the saddle point geometries, the original idea of the creation of the universe from `nothing' or an initial zero-size universe might have to be traded off in favor of other suitable initial boundary conditions that lead to stable saddle points \cite{di2019no,di2019noprescription,rajeev2021wave}. Even though we shall neither use the Hartle-Hawking nor the Vilenkin saddle points, we find that the wave function we derive bears more resemblance to the Vilenkin's tunneling wave function than it does to that of Hartle-Hawking's no-boundary proposal. This conclusion is surprisingly consistent with earlier studies of the de Sitter Swampland conjecture in the context of quantum cosmology \cite{brahma2020swampland,matsui2020swampland}.\par
A recently proposed swampland conjecture called the Trans-Planckian Censorship Conjecture (hereafter referred to only as TCC) posits that the expansion of the universe must be such that perturbation modes with sub-Planckian length scales should never leave the Hubble horizon and classicalize \cite{bedroya2020trans}. This restriction renders a long-lived de Sitter evolution nonviable. The universe's wave function computed for de Sitter spaces must reflect the above constraint posed by the TCC. Thus, we expect the wave function of the de Sitter universe to be heavily suppressed for conditions violating the trans-Planckian censorship.\par
This suppression cannot come from naively applying (semiclassical) quantum mechanics to General Relativity (GR) alone, as it is an expected quantum gravity effect. Thus, we propose modifying the Einstein-Hilbert action by using an additional boundary term with a complex part. Such modification can be thought of as an effective low-energy correction coming from the correct theory of quantum gravity, which encompasses phenomena such as the \textit{trans-Planckian censorship} (see, for example, works \cite{PhysRevLett.121.201301,PhysRevD.102.106023} that in different context deal with quantum gravity corrections to minisuperspace quantum cosmology). While the real part provides an unimportant phase factor to the wave function, the complex part, chosen judiciously, provides the expected exponential suppression. This \textit{postulated} modification being only a boundary term does not affect the classical dynamics. Therefore, while de Sitter spaces are allowed classically, the quantum mechanical amplitude of a universe evolving into a configuration that violates TCC is diminishingly small. \par
We shall see that the mathematical expression for the TCC constraint is best translated to our context if we choose to fix the initial and final Hubble rate at the boundaries. This can be done by including, apart from the \textit{postulated} TCC boundary term, additional covariant Robin boundary terms, which are proportional to the volume of the boundary hypersurfaces \cite{krishnan2017robin} (see \cite{di2019noprescription} for an example of using such a boundary condition in the context of quantum cosmology. Also, see \cite{chakraborty2017boundary} for a formal discussion on suitable boundary terms alternative to the Gibbons–Hawking–York boundary term while dealing with the action principle in GR). As the TCC term, by construction, does not alter the saddle point structure of the main action, our analysis is similar to that in \cite{di2019noprescription}, however, with the exception that we are required to fix the Hubble rate at both the initial and final boundaries.\par
Finally, we shall see that the two saddle points of the action considered here are real and lead to Lorentzian geometries describing the contraction of the universe from a given initial Hubble rate to the waist of the de Sitter hyperboloid and then expansion up to the given final Hubble rate. However, we shall find it useful to choose the initial Hubble rate to be zero, which then describes the geometry of only expanding universes that start from the waist of the de Sitter hyperboloid and expand till the final Hubble rate is reached. Moreover, the stability analysis for the perturbations around the two saddle points will help us decide their physical viability.\par
As an immediate consequence of modifying the universe's wave function, the TCC suppression factor predicts a characteristic value for the final Hubble rate at the end of inflation. We find that the factor leaves a significant imprint on the primordial power spectrum of the quantum fluctuations through scale non-invariance. In contrast, in the limit of the absence of quantum gravity corrections, the scale-invariant spectrum corresponding to the standard Bunch-Davies vacuum is recovered. However, as the scale invariance of the power spectrum is favored, deviations from such invariance should be limited, leading to the conclusion that the energy scale for inflation should be much below the Planck energy. This conclusion is consistent with the implications of TCC \cite{bedroya2020transinflation}.\par
We organize the paper as follows: In section \ref{sec:TCC}, we briefly recall the motivation and the mathematical statement for the Trans-Planckian Censorship Conjecture. In section \ref{sec:TCC_constraint_off-shell}, we analyze how the TCC constrains the off-shell geometries that contribute to the path integral sum. We find that the TCC imposes restrictions on the initial and final Hubble rates, which, in turn, motivates us to consider fixed Hubble rates as initial and final boundary conditions. With these boundary conditions, we explicitly compute the TCC modified wave function of the universe in the section \ref{sec:wavefunction}. In the following section \ref{sec:Hubble_rate_suppression}, we discuss the suppression of the Hubble rate at the end of inflation compared to its classically expected value as an immediate consequence of the modified wave function. Finally, in section \ref{sec:perturbation}, we deal with primordial gravitational wave perturbations against the background saddle point geometry, which lead to non-trivial scale dependence in the power spectrum, whereas when the quantum gravity effects are turned off, we recover the standard scale-invariant spectrum.

\section{Statement of the Trans-Planckian Censorship Conjecture}\label{sec:TCC}

The observation of the Cosmic Microwave Background (CMB) supports the idea of a homogeneous and isotropic universe at large scales. However, the existence of an incredible correlation between the patches in the sky that are causally disconnected, such as a high degree of uniformity of the CMB temperature \cite{aghanim2020planck} at large scales, poses a challenge in explaining how such an equilibrium came to be established despite the local physics being limited by causal horizons.\par
The inflationary paradigm offers an elegant explanation for the causal origin of the large-scale structures in the universe \cite{guth1981inflationary,linde1982new,starobinsky1980new}. In this framework, quantum vacuum fluctuations which are causally connected leave the Hubble horizon---the limiting distance below which length scale causal physics can establish equilibrium---subsequently get squeezed and classicalized before re-entering the horizon (at a later era when the inflation has stopped and has been replaced with the ``standard'' big-bang evolution) to form the large scale structure observed in the present-day CMB. The fluctuation modes leave the horizon during a phase of exponential expansion of the universe when the Hubble horizon shrinks in the comoving coordinates, while the comoving length scales of the modes remain fixed. It is known that if this phase of expansion lasted slightly longer than the period minimally required to explain the causal origin of observed homogeneity and isotropy in the large scale structure, certain length scales we see today can be traced back in time to have their origin in length scales smaller than the Planck length, a regime where the framework of quantum fields on curved classical background spacetime is supposed to breakdown. This problem in the inflationary paradigm is known as the \textit{trans-Planckian issue} \cite{martin2001trans}.\par
It is often argued that if particle production only occurs at super-Planckian length scales, then the effects of modified short-wavelength physics will not be visible in the CMB. Hence, there may not be a trans-Planckian issue. This line of thought assumes the trans-Planckian evolution to be essentially adiabatic. However, as we extrapolate our knowledge to sub-Planckian length scales, our conventional physics cannot be trusted anymore; moreover, we are unaware of the correct quantum gravitational physics appropriate for this regime, aptly referred to as the ``trans-Planckian zone of ignorance'' in the literature. In such trans-Planckian energy scales, non-adiabatic processes may occur, and particles may be created as a result. Then, as the physics operating at the scale of short-wavelength modes is different, a scale dependence or tilt in the power spectrum may be expected if these trans-Planckian modes get out of the horizon, classicalize, and re-enter the horizon at a later time (see, for example, \cite{PhysRevD.68.063513,brandenberger2013trans} and references therein). This is an issue because, on the one hand, the predictions of the inflationary paradigm should be robust against our ignorance of the short-wavelength physics; on the other hand, we do not know how to handle these trans-Planckian modes, and there are indications that consideration of `new' physics may affect the power spectrum.\par
Recently, it has been conjectured that in an effective low-energy field theory consistent with the quantum theory of gravity, the expansion of the universe should be such that this problem never arises, that is, the modes with sub-Planckian length never leave the Hubble horizon and classicalize \cite{bedroya2020trans}. Therefore, no perturbation mode with super-Hubble length scale can be traced back to sub-Planckian length scales at earlier times. The field theories that lead to potentials such that the expansion of the universe classicalizes those trans-Planckian modes are inconsistent and belong to the Swampland.\par
Mathematically, this conjecture can be formulated in terms of initial and final scale factors of the expansion phase, $a_i$ and $a_f$, and the final Hubble parameter $H_f$ as
\begin{align} \label{eq:TCC01} \frac{a_f}{a_i}\cdot \ell_{\rm Pl}<\frac{1}{H_f}\implies \frac{a_f}{a_i}<\frac{M_{\rm Pl}}{H_f}, \end{align}
that is, modes with sub-Planckian length scale do not cross the Hubble horizon ($H^{-1}(t)$) to classicalize and thus remain quantum. Here, $M_{\rm Pl}=\ell_{\rm Pl}^{-1}$ is the reduced Planck mass.\par
The time-reversal of the above statement corresponds to a restriction on the contracting phase of a universe. No cosmological contraction consistent with the quantum theory of gravity leads to perturbations with length scales larger than the Hubble horizon ($-1/H$), assuming a sub-Planckian length scale due to the cosmic evolution. Mathematically, the criterion reads
\begin{align} \label{eq:TCC02} \frac{a_i}{a_f} < - \frac{M_{\rm Pl}}{H_i}. \end{align}
Therefore, the TCC constrains long-lived de Sitter spaces, and we shall incorporate this restriction into the wave function of the universe.

\section{TCC constraint on de Sitter geometries} \label{sec:TCC_constraint_off-shell}
\begin{figure}
    \centering
    \includegraphics[width = 0.5 \textwidth]{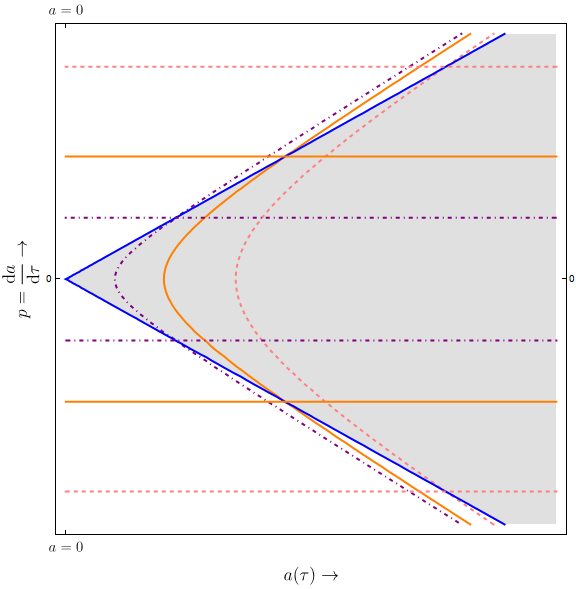}
    \caption{The solid (orange) curve represents the classical evolution of the universe in the $(p,a)$ plane when both the Friedmann equations (\ref{eq:Friedmann01}, \ref{eq:Friedmann02}) are satisfied. The dotted (pink) and dot-dashed (purple) curves represent off-shell geometries that follow the dynamical equation of motion but fail to satisfy the Hamiltonian constraint. The horizontal solid, dotted, and dot-dashed lines (TCC lines) provide the maximum (or minimum) $p$ value allowed by TCC (see, eq. (\ref{eq:parametrized_TCC})) for the respective geometries. The intersection points of the geometries and their respective TCC lines together form curves, which are the slopped (blue) straight lines representing the bounded region (gray) of evolution allowed by the TCC.}
    \label{fig:TCC}
\end{figure}
Our goal is to obtain the wave function of the universe, which is consistent with TCC---a conjecture that does not allow long-lived de Sitter evolution. The semiclassical quantum cosmology models remain oblivious to such a quantum gravity-motivated phenomenon. Therefore, to study this effect which arises from the physics of the super-Planckian energy scales, we must resort to a phenomenological approach in which we modify, by hand, certain assumptions in our low-energy theories hoping to have the effect in question be included in the `corrected' low-energy theories, by construction.\par
Therefore, we believe the path integral problem has to be modified in order for the TCC restrictions to reflect in the wave function. The reasoning is that as the correct theory of quantum gravity does not allow for long-lived de Sitter evolution, by suppressing the contribution coming from the geometries that violate such TCC constraint to the path integral sum, we shall be able to achieve a wave function that assigns lower probabilities for the universe's evolution into the configurations violating the TCC. One possible but not necessarily unique way to change the path integral problem to include the TCC constraint is to consider a modified principle for the path integral, wherein the paths (or geometries) that violate the TCC come into the path integral sum with an exponentially suppressed weightage. Such weighting can be achieved through adding appropriate complex terms to the classical Einstein-Hilbert action, as we have already alluded to before.\par
Notice that in the conventional path integral sum, all possible geometries contribute equally because there is no \textit{a priori} reason to distinguish the geometries from one another, i.e., the classical path or the extremum path has no special status, at least at the level of the defining the path integral. Suppose the path integral is modified to facilitate phenomenological inclusion of quantum gravity effects, as we propose to do here to include the effects of TCC. In that case, the simplest kind of modification should preserve as many of the properties of conventional physics as possible. Therefore, we have chosen not to break this equal treatment of all paths (or geometries) and apply the TCC restriction to both off-shell and on-shell geometries, even though the perturbations against the off-shell background geometries are not well understood. Therefore, it must be analyzed how the TCC restricts the off-shell geometries.\par
At this point, we would like to reference the work \cite{PhysRevLett.78.1854}, in which, in a different context, the quantum gravity effects were included by modifying the weightage of the possible paths that appear in the path integral sum. It is often argued that quantum gravity leads to a fundamental scale of length, the Planck length, the lowest possible physical distance up to which physical processes can occur. To include this effect in the path integral formalism, where the conventional version remains oblivious to the existence of the Planck length, the weightage of the paths contributing to the integral is modified such that the contribution of the paths in the integral with (Euclidean) path length smaller than the Planck length is exponentially suppressed. Notice that this modified weightage in the path integral sum restricts both on-shell and off-shell geometries within this framework. Our argument here is analogous. The key difference is that we achieve the suppression of the geometries violating TCC only by adding boundary terms to the action, which does not affect the underlying quantum mechanics.\par
Before proceeding to perform the path integral to compute the wave function of the TCC constrained universe, we must translate the conditions (\ref{eq:TCC01}) and (\ref{eq:TCC02}) such that these pose a meaningful constraint on the geometries involved in the path integral sum. As in the path integral, we sum over all possible geometries given particular boundary conditions; we need to figure out how the above TCC conditions restrict the off-shell geometries.\par
Let us consider the minisuperspace metric of a closed FLRW type universe
\begin{align} {\rm d}s^2 = - {\rm d} \tau^2 + a^2(\tau)\,{\rm d}\Omega_3^2, \end{align}
where we have denoted the scale factor and the metric on the 3-sphere as $a(\tau)$ and ${\rm d}\Omega_3^2$, respectively. In the absence of any matter fields, the evolution of the universe driven by a positive cosmological constant $\Lambda \equiv 3 H^2$ is given by the following Friedmann equations
\begin{align} \label{eq:Friedmann01} \frac{{\rm d}^2 a}{{\rm d}\tau^2} - H^2 a(\tau) = 0, \\ \label{eq:Friedmann02} \left(\frac{{\rm d}a}{{\rm d}\tau}\right)^2 + 1 - H^2 a^2(\tau) = 0. \end{align}
The classical evolution of the universe should be such that both the dynamical and constraint equations are satisfied. Now consider the off-shell geometries, which satisfy the dynamical Friedmann equation, but do not satisfy the Hamiltonian constraint. These geometries can be parameterized with a single parameter $\sigma$ as follows
\begin{align} \label{eq:off-shell_geometries} \left(\frac{{\rm d}a}{{\rm d}\tau}\right)^2 + 1 + \sigma - H^2 a^2(\tau)=0 \end{align}
For the on-shell geometry to satisfy both the dynamical and constraint equations, the parameter $\sigma$ should vanish. The minimum size or the size of the waist of these geometries is obtained by solving $\frac{{\rm d}a}{{\rm d}\tau} = 0$, which gives
\begin{align}\label{eq:waist} a_{\rm min} = \frac{\sqrt{1+\sigma}}{H}, \end{align}
where we must have $1+\sigma\geq0 \implies \sigma \geq -1$ for a real and positive solution for $a_{\rm min}$ when $\frac{{\rm d}a}{{\rm d}\tau} = 0$. As these geometries contribute to the path integral, we demand that these be constrained by the TCC criteria as well. As the different off-shell geometries have different minimum values for the scale factor, the bounds on their evolution by the TCC are supposed to be adjusted accordingly. Then, the parametrized TCC criteria for these off-shell geometries read
\begin{align}\label{eq:parametrized_TCC} \frac{a(\tau)}{a_{\rm min}} < \frac{M_{\rm Pl}}{H(\tau) } \implies \frac{{\rm d}a }{{\rm d} \tau} < \frac{M_{\rm Pl}\sqrt{1+\sigma}}{H}. \end{align}
Thus, given a parameter value for $\sigma$, the TCC condition imposes an upper bound for the value of $p=\frac{{\rm d}a}{{\rm d}\tau}$. Similarly, a lower bound comes from the statement of TCC corresponding to the contracting branches of these geometries. In general, the family of curves demarcating the boundary for the region of expansion allowed by the TCC for the respective family of curves for the off-shell geometries is given by the following equation
\begin{align}\label{eq:TCC_line} 1 + \sigma = \left(\frac{H}{M_{\rm Pl}} \frac{{\rm d}a}{{\rm d} \tau}\right)^2. \end{align}
For any allowed value of $\sigma$, the off-shell geometry is intersected by the TCC line(s) given by the above equation (\ref{eq:TCC_line}). The points of intersection of these two families of curves---the off-shell geometries and the respective TCC bound on these---form curves in the $p=\frac{{\rm d}a}{{\rm d}\tau}$ vs. $a$ plane, which demarcate the boundary of the region of evolution allowed by TCC for all possible off-shell geometries (see figure \ref{fig:TCC}). The equation for the curves formed by the intersection of the family of curves (\ref{eq:off-shell_geometries}) and (\ref{eq:TCC_line}) is obtained by eliminating $\sigma$ from these two equations and reads
\begin{align} p = \pm\frac{H}{\sqrt{1+\frac{H^2}{M_{\rm Pl}^2}}}a. \end{align}
Therefore, the TCC allowed region in the $(p,a)$ plane or the `phase plane' is bounded by two straight lines with slopes $\pm H/\sqrt{1+\frac{H^2}{M_{\rm Pl}^2}}$.\par
The TCC criterion is then translated into the following inequalities in the $(p,a)$ plane
\begin{align} \label{eq:TCC_boundary01} p(\tau) - \frac{H}{\sqrt{1+\frac{H^2}{M_{\rm Pl}^2}}} a (\tau) < 0,\\ \label{eq:TCC_boundary02} p (\tau ) + \frac{H}{\sqrt{1+\frac{H^2}{M_{\rm Pl}^2}}} a (\tau) > 0, \end{align}
for expanding and contracting branches of the geometries, respectively.
For any geometry with $p(\tau)\pm{H a(\tau)}/{\sqrt{1+\frac{H^2}{M_{\rm Pl}^2}}}$ evaluated at the initial or final boundaries (or at the initial and final times $\tau_{i}$ or $\tau_{f}$), if the conditions (\ref{eq:TCC_boundary01}) or (\ref{eq:TCC_boundary02}) are violated, then these geometries should receive an exponentially suppressing weight as discussed above.\par
Moreover, observe that the quantity $p(\tau)/a(\tau)$ is the Hubble rate at a particular time $\tau$. Then the TCC conditions imply constraints on the initial and final Hubble rates. With this observation, we see that the imposition of the TCC in the path integral approach is best handled if we choose to fix the Hubble rates at the initial and final boundary hypersurfaces.\par

\section{The wave function} \label{sec:wavefunction}
Schematically, computing the wave function $\Psi(H_f)$ in terms of the final Hubble rate in the path integral formalism amounts to performing the following sum
\begin{align}
    \Psi(H_f)=\sum_{\rm geometries} e^{iS},
\end{align}
where, $S$ is the action of the system. A suitable action for our context should have these three components: (a) the conventional Einstein-Hilbert action $S_{\text{E-H}}=\frac{1}{2\kappa}\int {\rm d}x^4\sqrt{-g}(R-6H^2)$; (b) suitable boundary terms $S_B$, which let us fix the initial and final Hubble rates; (c) an additional \textit{postulated} term $S_{\rm TCC}$, supposed to have its origin in a correct theory of quantum gravity, which has a complex part providing exponentially suppressing weight for the amplitudes violating conditions (\ref{eq:TCC_boundary01}) and (\ref{eq:TCC_boundary02}).\par
We choose to work in the minisuperspace with its metric, for convenience, being parametrized as follows \cite{halliwell1989steepest}:
\begin{align} {\rm d}s^2 = - \frac{N^2(t)}{q(t)}{\rm d}t^2 + q(t) {\rm d}\Omega_3^2, \end{align}
where, the new scale factor is $q=a^2$, and the time coordinate $t\in [0,1]$ is related to the cosmic time through ${\rm d}\tau=N(t){\rm d}t/\sqrt{q(t)}$ with $N(t)$ being the lapse function.\par
Then, the corresponding action for the path integral sum has the form
\begin{align} S = & {M_{\rm Pl}^{2}}{V_3}\int_0^1 {\rm d}t \left[\frac{3}{2N}q\ddot{q}+\frac{3}{4N}\dot{q}^2+3N(1-H^2q)\right] + S_{B} + S_{\rm TCC}, \end{align}
where we have chosen to work with the gauge condition $\dot{N}=0$, and overdot denotes derivative with respect to $t$. $V_3$ is the volume of the closed unit 3-sphere. The term $S_B$, which fixes initial and final Hubble rates, consists of covariant Robin boundary terms \cite{krishnan2017robin,chakraborty2017boundary} on the corresponding boundary hypersurfaces
\begin{align} S_B & = -\frac{M_{\rm Pl}^{2}}{\zeta} \int_{\partial\mathcal{M}_1} \sqrt{h} \, {\rm d}^3y - \frac{M_{\rm Pl}^{2}}{\xi} \int_{\partial\mathcal{M}_0} \sqrt{h} \, {\rm d}^3y \nonumber\\ & = -\frac{M_{\rm Pl}^{2}}{\zeta} q(1)^{\frac{3}{2}} V_3 - \frac{M_{\rm Pl}^{2}}{\xi} q(0)^{\frac{3}{2}} V_3, \end{align}
where $\partial\mathcal{M}_1$ and $\partial\mathcal{M}_0$ are the space-like boundary hypersurfaces at $t=1$ and $t=0$, respectively, with $h_{ij}$ being the induced metric on these hypersurfaces; and $\xi$ and $\zeta$ are two constants.\par
The variation of above action leads to the following
\begin{align} \label{eq:variation} {\delta S} = & {M_{\rm Pl}^{2}}{V_3}\int_0^1 {\rm d}t \left[\frac{3}{2N}\ddot{q}-3NH^2\right]\delta q  + \frac{3}{2} {M_{\rm Pl}^{2}}{V_3} q(1) \delta \left(\frac{\dot{q}(1)}{N} - \frac{2}{\zeta} \sqrt{q(1)}\right) \nonumber\\ & - \frac{3}{2} {M_{\rm Pl}^{2}}{V_3} q(0) \delta \left(\frac{\dot{q}(0)}{N} + \frac{2}{\xi} \sqrt{q(0)}\right)+\delta S_{\rm TCC}. \end{align}
In order for the variational problem to be consistent, we must impose the following Robin boundary conditions on the initial and final boundary hypersurfaces
\begin{align} \label{eq:boundary} \frac{\dot{q}(1)}{N} - \frac{2}{\zeta} \sqrt{q(1)} = 0, \quad \frac{\dot{q}(0)}{N} + \frac{2}{\xi} \sqrt{q(0)} = 0. \end{align}
Realizing that the Hubble rate $\frac{1}{a}\frac{{\rm d}a}{{\rm d}\tau}$ translates to $\frac{\dot{q}}{2N\sqrt{q}}$ in the new metric, we relate the constants $\xi$ and $\zeta$ with the initial and Hubble rates as \begin{align} \frac{1}{\zeta} = \frac{\dot{q}(1)}{2N\sqrt{q(1)}} = H_1, \quad \frac{1}{\xi} = - \frac{\dot{q}(0)}{2N\sqrt{q(0)}} = -H_0. \end{align}
Therefore, by imposing the Robin boundary conditions at both the boundaries, we have fixed the Hubble rates as desired.\par
Now, the wave function is formally obtained by performing the following path integral
\begin{align}
    \Psi(H_1)=\int {\rm d}H_0\int_{-\infty}^{\infty} {\rm d}N \int_{\frac{\dot{q}(0)}{2N\sqrt{q(0)}}=H_0}^{\frac{\dot{q}(1)}{2N\sqrt{q(1)}}=H_1} \mathcal{D}[q]e^{iS[q,N]} \Sigma(H_0),
\end{align}
where $\Sigma(H_0)$ is the universe's initial state yet to be chosen, and the integration over $H_0$ ranges up to all its possible values.\par
In the semiclassical limit, the $q$ integral corresponds to the dynamical equation of motion $\ddot{q}=2N^2H^2$ (since $\frac{\delta S}{\delta q}=0$ implies the EoM) and picks up the most contribution from geometries $\bar{q}(t)$ that satisfy this EoM along with the boundary conditions (\ref{eq:boundary}) but not necessarily the constraint equation, and gives the following result
\begin{align} \label{eq:lapse_integral}
    \Psi(H_1)\approx\int {\rm d}H_0 \int_{-\infty}^{\infty} {\rm d}N e^{iS[\bar{q},N]}\Sigma(H_0),
\end{align}
where we have ignored the prefactors coming from the quantum fluctuation integral in our semiclassical limit.\par
On the other hand, the lapse $N$ integral corresponds to the Hamiltonian constraint equation $\dot{q}^2=4N^2(H^2 q-1)$ (since $\frac{\delta S}{\delta N}=0$ implies the Hamiltonian constraint). Therefore, the integration over the lapse function amounts to summing over path integral amplitudes $e^{iS[\bar{q},N]}$, for all the geometries that satisfy the dynamical EoM but not necessarily the constraint equation, i.e., the geometries given by the equation (\ref{eq:off-shell_geometries}). As discussed earlier, these geometries have a specific region allowed by the TCC beyond which the amplitudes coming from these geometries in the path integral sum must receive exponentially decreasing weight. The exponential suppression should result from the complex part of the $S_{\rm TCC}$. Thus we must choose an appropriate form for this added term such that it does not alter the variational problem we are dealing with in (\ref{eq:variation}). Also, note that, due to the uncertainty principle, the off-shell paths or geometries that appear in the quantum fluctuation integral are highly irregular, and perturbations around these geometries are not well understood, in general. Thus the application of TCC to these geometries can be tricky. That is why we only deal with the TCC restriction during the lapse function integration, in which case the appearing off-shell geometries are de Sitter type geometries with different waist sizes (see, Eq. (\ref{eq:off-shell_geometries}) and (\ref{eq:waist})).\par
The geometries we are working with have fixed initial and final Hubble rates; therefore, these geometries start from a slopped straight line $p=H_0 a$ and end on the slopped straight line $p=H_1 a$ in the $(p,a)$ plane. If these lines happen to lie outside the bounded region allowed by the TCC (or outside the region demarcated by (\ref{eq:TCC_boundary01}) and (\ref{eq:TCC_boundary02})), then the quantum mechanical amplitude should be heavily suppressed; otherwise, the suppression should be relaxed. The conditions for TCC violation read
\begin{align} \frac{\dot{q}(1)}{2N\sqrt{q(1)}}=H_1 > \frac{H}{\sqrt{1+\frac{H^2}{M_{\rm Pl}^2}}},\\ \frac{\dot{q}(0)}{2N\sqrt{q(0)}}=H_0<-\frac{H}{\sqrt{1+\frac{H^2}{M_{\rm Pl}^2}}}.\end{align}
Then the suppression of geometries with the above TCC violating conditions can be achieved with a simple \textit{ansatz} for the complex part of the $S_{\rm TCC}$ term
\begin{align} S_{\rm TCC} & = i \frac{(H_1-H_0)}{2H}\sqrt{1+\frac{H^2}{M_{\rm Pl}^2}}, \end{align}
where the factor of $1/2$ has been included to account for the fact that probability distribution is defined as the (modulus) square of the amplitude. This is purely a boundary term and as $H_0$ and $H_1$ are fixed quantities on the boundary, the variation of $S_{\rm TCC}$ vanishes. Thus the addition of this term does not affect the variational problem keeping the dynamical equation unaffected as we explain in the following.\par
Notice that we have fixed the initial and final Hubble rates as boundary conditions. This is achieved by means of adding covariant Robin boundary terms. Now, the TCC boundary terms are linear functions of initial and final Hubble rates. As the Hubble rates at the boundaries are fixed, the variation of the TCC boundary terms must vanish. See the total variation of the action in the equation (\ref{eq:variation}) above, which shows all the relevant terms when the action is varied. Then, if we impose the boundary conditions that the Hubble rates at both the boundaries are constants, the variations of all the boundary terms, including the variation of the TCC boundary terms $\delta S_{\rm TCC}$ vanish, leaving only the non-boundary variation term associated with $\delta q$. This leads to the standard equation of motion with boundary conditions where Hubble rates at both ends are fixed. Therefore, adding the TCC boundary terms in this specific instance does not change the variational problem. However, this may not be true in general. In a generic situation, adding new boundary terms may as well require one to impose a different set of boundary conditions such that the variation of the boundary terms vanish altogether and the variational problem remains consistent.\par
Moreover, it does not explicitly depend upon $N$, leaving the constraint equation intact. For the purpose of the lapse integration, this term is a constant. In order for the dynamical and constraint equations to not get affected, the real part of the $S_{\rm TCC}$ must also be a constant, and hence it only contributes to an overall unimportant phase to the wave function. Even though the addition of the TCC term, by construction, does not affect the lapse integration, it leaves an imprint on the primordial power spectrum, as will be discussed in section \ref{sec:perturbation}.\par
Moreover, notice that due to simple nature of our \textit{ansatz} for the TCC correction term, it in fact affects neither the quantum fluctuation integral in $\int \mathcal{D}[q]e^{iS}$ (as $H_1$ and $H_0$ are fixed on the boundary) nor affects the lapse integration $\int {\rm d}N e^{iS(N)}$ (as the TCC term has no explicit $N$ dependency). This fact allows the term bearing $S_{\rm TCC}$ in the exponent to be factored out of the $q$ and $N$ integrals entirely, and we can express the wave function as $\Psi(H_1)=\int {\rm d}H_0 F_{\rm TCC}(H_1,H_0)\psi(H_1,H_0)\Sigma(H_0)$, where $F_{\rm TCC}$ is equal to $e^{iS_{\rm TCC}}$ and $\psi(H_1,H_0)$ is the wave function calculated from the conventional Einstein-Hilbert action and the covariant Robin boundary terms. Therefore, the quantity $\psi(H_1,H_0)$ should satisfy all the properties expected of a quantum mechanical amplitude in the context of conventional quantum cosmology. We are calling the transition amplitude $\psi(H_1,H_0)$ as a wave function because the full range of integration for the lapse function $(-\infty,\infty)$ is known to produce amplitudes that satisfy the quantum mechanical version of the Hamiltonian constraint equation, also known as the Wheeler-DeWitt equation \cite{halliwell1989steepest}.\par
To evaluate $S[\bar{q},N]$, we have to determine the solution $\bar{q}(t)$ to the classical equation of motion $\ddot{q} = 2 N^2 H^2$ such that the Robin boundary conditions (\ref{eq:boundary}) are satisfied. The solution $\bar{q}(t)$ reads
\begin{align} \bar{q}(t) = H^2 {N}^2
   t^2+\frac{2 {N}^2 \left(\sqrt{H^2 \zeta^2-1} \sqrt{H^2
   \xi ^2-1} - H^2 \zeta^2 +1\right)}{\zeta ^2-\xi ^2} t \nonumber\\+ \frac{\xi ^2 {N}^2 \left(H^2 \zeta ^2 - 1\right)
   \left(\sqrt{H^2 \zeta ^2 -1}-\sqrt{H^2 \xi
   ^2-1}\right)^2}{\left(\zeta ^2-\xi ^2\right)^2}.
\end{align}

The full action
\begin{align} \frac{M_{\rm Pl}^{-2} S}{V_3} = & \int_0^1 {\rm d}t \left[\frac{3}{2N}q\ddot{q}+\frac{3}{4N}\dot{q}^2+3N(1-H^2q)\right] - \left(\frac{1}{\zeta} q(1)^{\frac{3}{2}} + \frac{1}{\xi} q(0)^{\frac{3}{2}}\right) \nonumber\\ & +i\frac{M_{\rm Pl}^{-2}}{2V_3} \frac{(\xi + \zeta)}{\xi \zeta H}\sqrt{1+\frac{H^2}{M_{\rm Pl}^2}}, \end{align}
then, can be evaluated at the classical solution $\bar{q}(t)$ and the result is
\begin{align} \frac{M_{\rm Pl}^{-2} S(N)}{V_3} = & 3 N - \frac{N^3 A^2(\xi) B^2(\zeta) \left(B(\zeta)-A(\xi)\right)^2}{\left(\zeta ^2-\xi
   ^2\right)^2} + i \frac{M_{\rm Pl}^{-2}}{2V_3} \frac{(\xi + \zeta)}{\xi \zeta H}\sqrt{1+\frac{H^2}{M_{\rm Pl}^2}}, \end{align}
here, we have defined
\begin{align}
A(\xi) = \sqrt{H^2\xi^2 -1 }, \quad B(\zeta) = \sqrt{H^2\zeta^2 -1 }.
\end{align}
The lapse integral of the kind (\ref{eq:lapse_integral}) that sums up oscillatory functions $e^{\frac{i}{\hbar}S(N)}$ is not absolutely convergent, but in simple cases can be conditionally convergent. To perform this integration (see \cite{feldbrugge2017lorentzian}), one complexifies the lapse function $N$ and deforms the real line of integration into the complex plane in accord with Picard-Lefschetz theory such that the conditional convergence is made into absolute convergence. The contour of integration is chosen to align with the Lefschetz thimbles or steepest descent contours along which the real part of the action ${\rm Re}[S(N)]$ freezes to a constant---thus, the oscillatory function does not oscillate anymore---and the complex part of ${\rm Im}[S(N)]$ is such that the value of $e^{-\frac{1}{\hbar}{\rm Im}[S(N)]}$ decreases rapidly downward from the saddle points. In the semiclassical limit, $\hbar\to0$, the integration is approximated by a sum of the integrand values at the saddles points.\par
It is straightforward to determine the saddle points $N_{s\pm}$ corresponding to the action $S[\bar{q},N]$ in the complex $N$ plane by solving the equation $\frac{\partial S[\bar{q},N]}{\partial N}=0$ for $N$, and the solutions read
\begin{align} N_{s\pm} = \pm \frac{\zeta ^2-\xi
   ^2}{A(\xi) B(\zeta)
   \left(B(\zeta)-A(\xi)\right)}. \end{align}
As we have considered closed slicing of the de Sitter space, we must have $H_0,H_1\leq H$, or equivalently $H^2\zeta^2-1\geq 0$ and $H^2\xi^2-1\geq 0$, and as a result, the saddle points are real and correspond to Lorentzian geometries. Given these saddle points, it is easy to evaluate the action at the saddle points to be
\begin{align} S(N_{s\pm}) = & \pm \frac{2V_3}{M_{\rm Pl}^{-2}} \frac{(H_1^2 -H_0^2)\left({H^2-H_1^2}\right)^{-\frac{1}{2}}\left({H^2-H_0^2}\right)^{-\frac{1}{2}}}{\left(H_1 \sqrt{H^2-H_0^2} - H_0 \sqrt{H^2-H_1^2}\right)} +i \frac{(H_1 - H_0)}{2H}\sqrt{1+\frac{H^2}{M_{\rm Pl}^2}}.\end{align}
Then the wave function $\psi$ along with the factor $F_{\rm TCC}$ can be approximated as
\begin{align} F_{\rm TCC}\times\psi \simeq e^{i S(N_{s+})/\hbar} + e^{i S(N_{s -})/\hbar}, \end{align}
leading to the cosine function of $|S(N_{s\pm})|/\hbar$. Even though this wave function would be mathematically correct, whether this represents our physical universe depends on the behavior of perturbations around the background geometry corresponding to the saddle points. If the perturbations around any of these saddle points grow uncontrollably, i.e., lead to instability, then that saddle point cannot be considered physical. As we shall see in section \ref{sec:perturbation}, the perturbations around both the saddle points cannot be made well-behaved simultaneously; that is, if the perturbation around one saddle point is stable, then around the other saddle point, it grows uncontrollably. Therefore, we must choose only one saddle point among $N_{s\pm}$.\par
\begin{figure}
    \centering
    \includegraphics[width=0.5\textwidth]{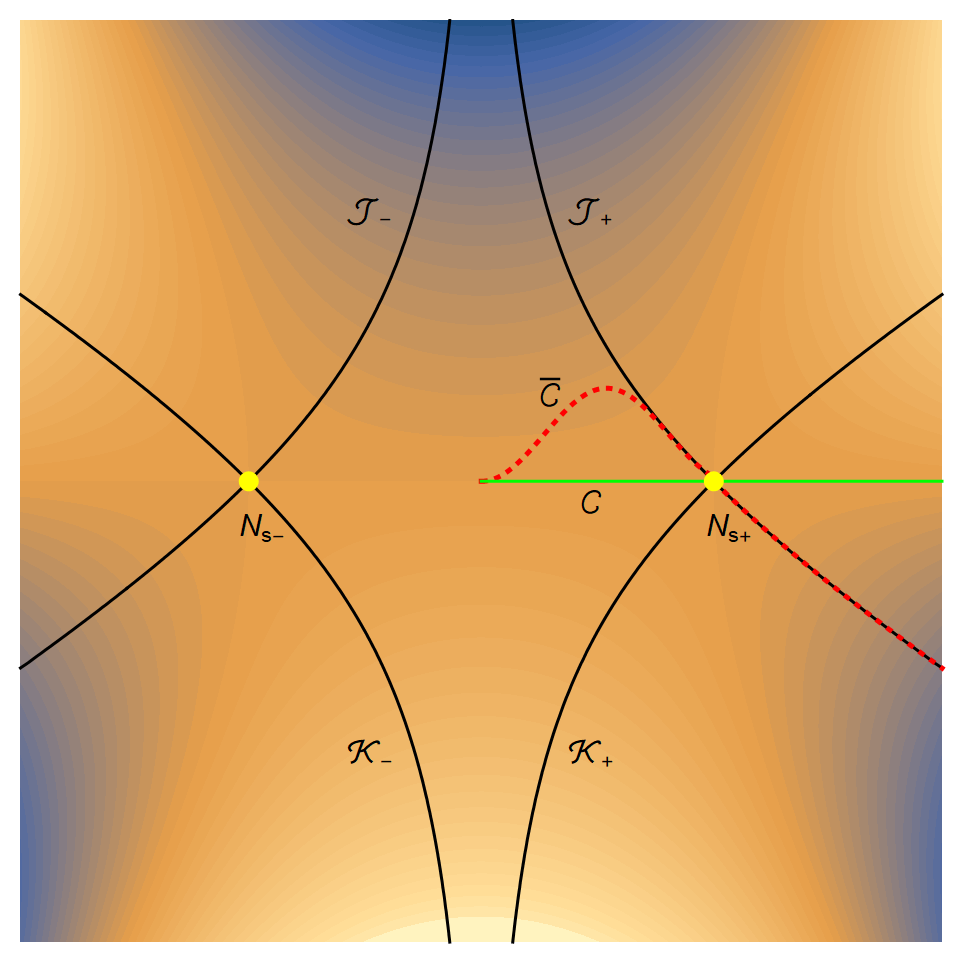}
    \caption{We have contour-plotted ${\rm Re}[iS(N)]$ in the complex $N$ plane. The bluer (darker) regions are the regions of absolute convergence of the integral, whereas the yellower (lighter) regions are the regions of divergence. The (yellow) dots $N_{s+}$ and $N_{s-}$ are the two saddle points. The solid (black) curves $\mathcal{J}_{\pm}$ are the steepest descent contours or Lefschetz thimbles. On the other hand, the solid (black) curves $\mathcal{K}_{\pm}$ are the steepest ascent flow lines. The original half-infinite $(0,\infty)$ integration contour $\mathcal{C}$, represented by solid (green) line is deformed into a new contour $\bar{\mathcal{C}}$, represented by a (red) dashed curve, such that it runs through the Lefschetz thimble $\mathcal{J}_{+}$ near the saddle point $N_{s+}$ to enable the saddle point approximation and then runs along $\mathcal{J}_{+}$ to infinity to ensure absolute convergence.}
    \label{fig:Picard-Lefschetz}
\end{figure}

This issue is averted by limiting the range for the lapse integration to $0<N<\infty$, which is known to produce the quantum mechanical amplitude for the universe to evolve to a state with Hubble rate $H_1$ starting from an initial state with $H_0$, that is
\begin{align}
    G[H_1,H_0]&=\int_{0^{+}}^{\infty} {\rm d}N \int_{\frac{\dot{q}(0)}{2N\sqrt{q(0)}}=H_0}^{\frac{\dot{q}(1)}{2N\sqrt{q(1)}}=H_1} \mathcal{D}[q]e^{i\tilde{S}[q,N]},\nonumber\\
    &\approx\int_{0^{+}}^{\infty} {\rm d}N e^{i\tilde{S}[\bar{q},N]}
\end{align}
where the action $\tilde{S}=S_{\text{E-H}}+S_{B}$ excludes the TCC term. The path integral $\int_{H_0}^{H_1} \mathcal{D}[q]e^{i\tilde{S}[q,N]}$ is the quantum mechanical propagator $G[H_1,H_0;N]$ for the universe that evolves from $H_0$ to $H_1$ in the proper time $N$ \cite{halliwell1989steepest}. The range of integration for the lapse function, $0 < N < \infty$ automatically chooses the saddle point $N_{s+}$, see figure \ref{fig:Picard-Lefschetz}.\par
Thus, in the semi-classical limit, the amplitude is approximated to
\begin{align} G [H_1,H_0] \simeq & \exp\left({i\frac{2V_3}{\hbar M_{\rm Pl}^{-2}} \frac{(H_1^2 -H_0^2)\left({H^2-H_1^2}\right)^{-\frac{1}{2}}\left({H^2-H_0^2}\right)^{-\frac{1}{2}}}{\left(H_1 \sqrt{H^2-H_0^2} - H_0 \sqrt{H^2-H_1^2}\right)}}\right). \end{align}
We note that this amplitude constitutes a Green's function for an appropriate Hamiltonian which has the usual operator form $\mathcal{H}\to i \frac{\delta}{\delta N}$, that is, the action of the Hamiltonian operator on it the returns Dirac delta function,
\begin{align}
    \mathcal{H} G[H_1,H_0] & = i\int_0^\infty {\rm d}N \frac{{\rm d}}{{\rm d}N}G[H_1,H_0;N]\nonumber\\
    &= iG[H_1,H_0;\infty]-iG[H_1,H_0;0]\nonumber\\
    &=-i\delta(H_1-H_0),
\end{align}
where we have assumed that the quantum mechanical propagators vanish in the limit $N\to\infty$. As a result, for the choice $H_0=0$ on the initial boundary hypersurface, which can be set by choosing the initial state to be $\Sigma(H_0)=\delta(H_0)$, and with the assumption that $H_1>0$, which is appropriate given that we observe an expanding universe, the quantum mechanical amplitude satisfies the Hamiltonian constraint equation $\mathcal{H}G[H_1,0]=0$ expected of a wave function. Therefore, the quantity $G[H_1,0]$ with $H_1>0$ can be identified with the conventional wave function of the universe $\psi(H_1)$ (see for a comparison \cite{vilenkin2018tunneling}). The choice to set $H_0$ to zero (and $H_1>0$) implies we are only dealing with geometries $\bar{q}(t)$ that describe expanding universe, excluding the contracting branches or the cases of de Sitter bounce. Then the path integral can be interpreted as a sum over only inflating universes. With this initial boundary condition, we have the total wave function of the universe including the TCC factor, i.e., the total wave function $\Psi(H_1)=F_{\rm TCC}(H_1)\times\psi(H_1)$, as
\begin{align}\label{eq:TCC_wavefunction} \Psi (H_1) \propto  & \exp\left({i\frac{2V_3}{M_{\rm Pl}^{-2}} \frac{H_1}{H^2\sqrt{H^2-H_1^2}}}\right) \exp\left({ - \frac{H_1}{2H}\sqrt{1+\frac{H^2}{M_{\rm Pl}^2}}}\right). \end{align}
Due to the presence of the exponential suppression factor, the wave function for the de Sitter universe no longer remains oblivious to the short-lived nature of the de Sitter spaces. Moreover, as we have included only a single saddle point in the path integral sum, the wave function is complex-valued, oscillatory, and contains a negative weight factor, akin to Vilenkin's tunneling wave function. This feature aligns with the results from recent studies that show the tunneling wave function might be preferable over the no-boundary wave function in the context of other Swampland conjectures \cite{brahma2020swampland,matsui2020swampland}.\par
In general, the state $\Psi(H_1)$ presents us with the quantum amplitude of a universe evolving into a geometric configuration such that it has a final Hubble rate $H_1$. Nevertheless, this knowledge of the final hypersurface cannot be extrapolated back into time to reveal any definite information concerning the initial configuration of the universe. This lack of knowledge of the universe's initial state invited several proposals, such as Hartle-Hawking's ``no-boundary proposal,'' Vilenkin's ``tunneling proposal'' \textit{et cetera}. In both of these proposals, the universe originates from `nothing' or zero sizes, \textit{i.e.}, $\bar{q}(0)=0$. For our case, we see from equation (\ref{eq:off-shell_geometries}) that any geometry that has $a(0)=0$ must have $p(0)=\pm\sqrt{-1-\sigma}$ with $\sigma\leq-1$. However, this class of geometries violate the TCC (i.e., lie outside the region bounded by (\ref{eq:TCC_boundary01}) and (\ref{eq:TCC_boundary02})), and hence we can not consider $\bar{q}(0)=0$ as a viable initial boundary condition while dealing with TCC. Therefore, we do not have the option to use either the no boundary or the tunneling proposal, and the choice $H_0=0$ should be thought of as a substitute for the initial boundary condition for the universe, leading to a path integral sum over only inflating universes (compare this with \cite{rajeev2021wave}, wherein a sum over all \textit{eventually} inflating geometries has been considered).\par
As the geometries $\bar{q}(t)$ only have $N^2$ dependency, both the saddle points $N_{s\pm}$ lead to the same geometry. With the condition $H_0=0$, the saddle point geometries $\bar{q}_{s}(t)$ has the following form
\begin{align} \bar{q}_s(t) &= t^2\frac{H_1^2}{H^2 \left(H^2-H_1^2\right)}+\frac{1}{H^2}, \end{align}
which describe the evolution of a universe starting with a initial size $1/H$ at $t=0$ (as the scale factor $a(0)=\sqrt{\bar{q}_s(0)}=1/H$) up to a configuration with Hubble rate $H_1$ at $t=1$.

\section{Suppression of Hubble rate} \label{sec:Hubble_rate_suppression}
An immediate consequence of the wave function (\ref{eq:TCC_wavefunction}) is that the quantum gravitational prediction for the value of the final Hubble rate of the universe at the end of inflation is suppressed in comparison with its classical expectation. To see this consider the probability of the universe evolving to a configuration with Hubble rate $H_1$
\begin{align}\label{eq:probability}
    \left|\Psi(H_1)\right|^2 \propto \exp \left( - \frac{H_1}{H}\sqrt{1+\frac{H^2}{M_{\rm Pl}^2}}\right),
\end{align}
where due to the modulus square, the oscillatory part drops out. In the context of quantum cosmology, it is only meaningful to define relative probabilities between two quantum states of the universe. Thus we can choose a fiducial quantum state with respect to which the relative probability of attaining all other states has to be defined, like $|\Psi(H_1)|^2/|\Psi(H_{\rm fiducial})|^2$. In our case, without loss of any generality, this fiducial state can be taken as the initial state of the universe, leading to the expression (\ref{eq:probability}).\par
In the classical evolution of the de Sitter universe, the final Hubble rate $H_{\rm classical}$ asymptotically approaches the value $H$ when the expansion of the universe is continued for a long time or in the large scale factor limit (as evident from equation (\ref{eq:Friedmann02})). However, in the presence of trans-Planckian censorship, the de Sitter spaces do not live long; as a result, the final Hubble rate at the end of inflation assumes a smaller value. This is evident in (\ref{eq:probability}), since the probability of the universe evolving up to Hubble rate $H_1$ quickly diminishes for $H_1>H/\sqrt{1+H^2/M_{\rm Pl}^2}$. Therefore, the universe is likely to assume the following characteristic Hubble rate at the end of inflation
\begin{align}
    H_{\rm quantum}\sim \frac{H}{\sqrt{1+\frac{H^2}{M_{\rm Pl}^2}}}.
\end{align}

Consider the similarity of this argument to the case for Yukawa potential, in which the potential having a form like $e^{-mr}/r$ leads to the conclusion that the range of interaction has an associated characteristic distance determined by $\sim m^{-1}$.\par
This information regarding $H_{\rm classical}$ is also included in the quantum gravitational prediction for the final Hubble rate after the de Sitter expansion. To see this, consider the limit in which quantum gravitational effects are unimportant, i.e., $M_{\rm Pl}\to\infty$. In this limit, $H_{\rm quantum}$ approaches the classically expected value $H$.\par
Realizing that the configuration of the universe at the end of the inflation provides the initial conditions for the standard big bang cosmology, this quantum gravitational suppression of the Hubble rate ($H_{\rm quantum}<H_{\rm classical}$) in the early universe is likely to leave an imprint on the primordial power spectrum as we shall see in the next section.

\section{Quantum perturbations against the background} \label{sec:perturbation}
As there is no matter in our system we can only have gravitational wave perturbations. The second order action for such perturbations against the background saddle point geometry describing only a single mode (labelled by $l$) with fixed polarization is given by \cite{PhysRevLett.119.171301,PhysRevD.97.023509}
\begin{align}\label{eq:perturbation_action} S^{(2)}_{l}[\phi,\bar{q}_{s},N_{s\pm}] = \frac{V_3}{2}\int_{0}^{1} {\rm d}t \, N_{s\pm} \left( \bar{q}^2_{s} \frac{\dot{\phi}^2}{N^2_{s\pm}} - l (l + 2) \phi^2\right). \end{align}
We shall be ignoring backreaction on the geometry from the perturbations in our analysis. In the semiclassical limit, the total wave function can be assumed to be separable into the wave functions for the background and perturbations, i.e., $\tilde{\Psi}_{\rm tot} (H_1,\phi_1) = \Psi(H_1)\chi(\phi_1),$ where we have derived $\Psi(H_1)$ earlier and the perturbation wave function $\chi(\phi_1)$ is defined as the following
\begin{align}
    \chi(\phi_1)=\int_{-\infty}^{\infty} {\rm d} \phi_0 \int_{\phi(0) = \phi_0}^{\phi(1) = \phi_1} \mathcal{D} [\phi] e^{i S_l^{(2)} [\phi, \bar{q}_s, N_{s\pm}]} \chi_0(\phi_0),
\end{align}
where $\chi_0(\phi_0)$ is the initial wave function pertaining to the gravitational wave perturbation, and we shall choose it to be a coherent state in the following. The equation of motion corresponding to the above action reads
\begin{align}\label{eq:perturbation_eom} \ddot{\phi} + 2 \frac{\dot{\bar{q}}_s}{\bar{q}_s} \dot{\phi} + \frac{N^2_{s\pm}}{\bar{q}^2_s} l ( l + 2 ) \phi = 0, \end{align}
which resembles the equation of motion for a time-depended Harmonic oscillator with the following definitions for the mass and the angular frequency parameters
\begin{align} m_{s\pm}(t) = \frac{V_3}{N_{s\pm}} \bar{q}_s^2 (t), \quad \omega_l^2 (t) = \frac{N_{s\pm}^2 l (l+2)}{\bar{q}_s^2 (t)}. \end{align}

The perturbation action (\ref{eq:perturbation_action}) can be re-expressed employing integration by parts as
\begin{align} S^{(2)}_l = & \frac{V_3}{2}\int_{0}^{1} {\rm d}t \Bigg[ -\frac{\bar{q}^2_s}{N_{s\pm}}\phi\left(\ddot{\phi}+\frac{2\dot{\bar{q}}_s}{\bar{q}_s}\dot{\phi} + \frac{N^2_{s\pm}}{\bar{q}^2_s} l ( l + 2 ) \phi\right) +\frac{{\rm d}}{{\rm d}t}\left(\frac{\bar{q}_s^2}{N_{s\pm}}\phi \dot{\phi}\right) \Bigg]. \end{align}
When $\phi(t)=\phi_1\frac{F_l(t)}{F_l(1)}$ is a solution for the equation of motion (\ref{eq:perturbation_eom}), the classical action reduces to the following simplified form
\begin{align} S^{(2)}_l = & \frac{V_3}{2}\int_{0}^{1} {\rm d}t \frac{{\rm d}}{{\rm d}t}\left(\frac{\bar{q}_s^2}{N_{s\pm}}\phi \dot{\phi}\right) \nonumber \\ = & \frac{m_{s\pm}(1)}{2}\frac{\dot{F}_l(1)}{F_l(1)}\phi_1^2 - \frac{\bar{q}_s^2}{N_{s\pm}}\phi(0) \dot{\phi}(0). \end{align}

Choosing the initial wave function of the gravitational wave perturbation to be a coherent state
\begin{align}
    \chi_0(\phi_0)\propto\exp\left(-\frac{m(0)\omega_l(0)}{2}{\phi^2(0)}+i\pi(0)\phi(0)\right),
\end{align}
where $\pi=\frac{\bar{q}_{s}^2\dot{\phi}}{N_{s\pm}}$ is the conjugate momentum associated with $\phi$, reduces the perturbation wave function to the following form
\begin{align}\label{eq:perturbation_wavefunction}
    \chi(\phi_1) \approx \mathcal{N}_l\exp\left[i\frac{m_{s\pm}(1)}{2}\frac{\dot{F}_l(1)}{F_l(1)}\phi_1^2\right], 
\end{align}
where $\mathcal{N}_l$ is a suitable normalization factor. See, for example, \cite{vilenkin2018tunneling}, where a coherent initial state has been used.

We recall that the function $F_l(t)$ satisfies the equation
\begin{align}\label{eq:function_equation}
    \ddot{F}_l + \frac{\dot{m}_{s\pm}}{m_{s\pm}}\dot{F}_l + \omega_l^2 F_l = 0.
\end{align}
As $\dot{m}_{s\pm}/m_{s\pm}$ and $\omega^2_l$ have the same values for both the saddle points $N_{s\pm}$, respectively, the function $F_l(t)$ is independent of choice of the saddle points. The two linearly independent solutions of the above equation are
\begin{align} f_l(t),g_l(t) = & \frac{1}{\sqrt{\bar{q}_s}} \left[ \frac{t - \delta}{t - \gamma}\right]^{\pm \frac{l+1}{2}} \Big\{ \left[1 \mp (l+1) \right]\left(\gamma - \delta\right) + 2 (t - \gamma) \Big\}, \end{align}
here, $\gamma$, and $\delta$ are the solutions to the equation $\bar{q}_{s}(t)=0$.\par
Now, the two roots for the equation $\bar{q}_s(t)=0$ are
\begin{align}
    \gamma = i\frac{\sqrt{H^2-H_1^2}}{H_1},\quad \delta = -i\frac{\sqrt{H^2-H_1^2}}{H_1},
\end{align}
which are independent of the choice of the saddle points as well. A general solution $F_l(t)$ to (\ref{eq:function_equation}) is a linear combination of the independent solutions, that is, $F_l=\alpha f_l+\beta g_l$, where $\alpha,\beta\in\mathbb{C}$. As both the solutions, $f_l$, $g_l$ are regular in the interval $t\in[0,1]$, we have to be open to the possibility of both of these modes contributing. Additional requirements need to be invoked to determine these parameters.\par
One way to determine the parameters $(\alpha,\beta)$ is to demand that in the limit $\ell_{\rm Pl}\to 0$ (or equivalently $M_{\rm Pl}\to\infty$), i.e., when the quantum gravitational corrections are unimportant, the profile of the perturbation matches precisely with that of arising from the Bunch-Davies vacuum which is predicted from inflationary paradigm. This power spectrum has also been computed within the framework of LoQC in previous studies, for example, in the context of the no-boundary wave function (see for example \cite{di2019no}).\par
For the saddle point $N_{s+}$, and for the choice $(\alpha = 0,\beta = 1)$, the argument of the exponential in (\ref{eq:perturbation_wavefunction}) has the following form
\begin{align}
    -\frac{i l (l+2) V_3}{2 \sqrt{H^2-H_1^2} \left(H_1+i (l+1)\sqrt{H^2 - H_1^2}\right)}\phi_1^2.
\end{align}
Evaluating this expression for the characteristic value of the Hubble rate $H_1\sim H/\sqrt{1+H^2/M_{\rm Pl}^2}$, which is the likely value of the Hubble rate at the end of inflation, we get
\begin{align}\label{eq:exp_argument}
    -\frac{l (l+2) V_3 \sqrt{1+\frac{H^2}{M_{\rm Pl}^2}} M_{\rm Pl}^2\left(1+\frac{H^2}{M_{\rm Pl}^2}\right)}{2 H^3 \left(H (l+1) \sqrt{1+\frac{H^2}{M_{\rm Pl}^2}}-i M_{\rm Pl}\sqrt{1+\frac{H^2}{M_{\rm Pl}^2}}\right)}\phi_1^2.
\end{align}
In the limit, $M_{\rm Pl}\to\infty$, when the quantum gravity corrections are unimportant, the above expression reduces to a familiar form
\begin{align}
    -\frac{l (l+1) (l+2) V_3}{2 H^2}\phi_1^2-\frac{i l (l+2) M_{\rm Pl} V_3}{2 H^3}\phi_1^2+\mathcal{O}\left[\frac{1}{M_{\rm{Pl}}}\right],
\end{align}
which corresponds to the perturbations being described by a Gaussian distribution and leads to a scale-invariant power spectrum expected in the case of a Bunch-Davies vacuum state for the perturbations.\par
However, when we consider the saddle point $N_{s-}$ with the same choices for the parameters as before $(\alpha=0,\beta=1)$ similar analysis reveals that the perturbations are described by an \textit{inverse} Gaussian distribution and hence grow uncontrollably. For the saddle point $N_{s-}$, in order to achieve the Bunch-Davies vacuum in the $\ell_{\rm pl} \to 0$ limit for a universe that has the quantum characteristic Hubble rate at the end of inflation, one must choose the parameter values $(\alpha=1,\beta=0)$. The fact that the solutions $f_l,g_l$ are linearly independent renders these choices mutually exclusive. Therefore, if we demand that quantum gravity corrected results be consistent with the standard form when such corrections are absent, then the perturbations around both the saddle points cannot be stable simultaneously, and only one of those two corresponds to a physical universe. Thus, we have allowed only one of the saddle points to contribute to the path integral, which then led to a complex-valued wave function bearing a resemblance to the Vilenkin's tunneling wave function.\par
Keeping the lowest order of quantum gravity correction in the expression (\ref{eq:exp_argument}) alive, we arrive at the following wave function for the gravitational wave perturbations
\begin{align}
    \chi(\phi)\simeq e^{-\frac{V_3}{2H^2}l(l+1)(l+2)\phi_1^2+\frac{V_3}{2H^2}l^2(l+1)(l+2)^2\frac{H^2}{M_{\rm Pl}^2}\phi_1^2}
    \times \text{phase}.
\end{align}
The first term in the exponent corresponds to the scale-invariant power spectrum predicted from the inflationary paradigm and varies as $\sim l^{-3}$ in the large $l$ limit, while the next term introduces scale non-invariant behavior varying as $\sim l^{-5}$. As the scale invariance of the power spectrum is a robust prediction of the inflationary paradigm and has strong motivations, deviation from such behavior must be limited in scope, which implies $H\ll M_{\rm Pl}$. This also implies the scale of inflation should be much below the Planck scale, that is, $V\ll M_{\rm Pl}$, where $V$ is the potential that drives inflation. This qualitative conclusion is in agreement with \cite{bedroya2020trans,brandenberger2021trans}.

\section{Conclusion}
The inflationary paradigm provides an explanation for the causal origin of the large-scale structure from primordial quantum vacuum fluctuations. However, inflation lasting for a long enough duration implies modes of certain length scales observed today, if traced back in time, had their origin from length scales below the Planck length during inflation. However, it has been conjectured that this \textit{trans-Planckian issue} never arises within the framework of an effective low energy theory that is consistent with the correct UV complete quantum theory of gravitation. Effective quantum field theories that produce results inconsistent with the theory of quantum gravity are expected to belong to the Swampland.\par
To avoid the \textit{trans-Planckian issue} one requires that the Hubble horizon never shrinks below the Planck length during the exponential phase of expansion to let trans-Planckian modes leave the horizon and classicalize. Therefore, the Trans-Planckian Censorship Conjecture imposes a constraint on the duration of inflation. Whereas in the inflationary paradigm, one utilizes the framework of quantum field theory on classical curved backgrounds, in the framework of quantum cosmology, one attempts to deal with a quantum background as well. In this respect, several proposals for the wave function of a de Sitter toy universe were proposed, such as the Hartle-Hawking's or the Vilenkin's proposal, \textit{et cetera}, and subsequent works also extended the scenario to include more realistic potentials that drive the inflation \cite{garay1991path,jonas2022revisiting}. However, these proposed semiclassical wave functions do not take into consideration the restriction imposed on the duration of inflation by Swampland conjectures. This motivated us to modify the wave function of a de Sitter universe to include the constraint coming from TCC so that the wave function more closely corresponds to an EFT belonging to the Landscape as opposed to the Swampland.\par
To modify the wave function, we demanded that the TCC constrains all the possible off-shell geometries that contribute to the lapse integration. This demand gave rise to a bounded region in the `phase space' of de Sitter evolution or the $(p,a)$ plane, inside which the evolution of the universe is supposed to be allowed by the TCC. Moreover, we found that the TCC constraint translates to restriction on the initial and final Hubble parameter value in the $(p,a)$ plane. Thus, we chose to set up the path integral problem with the boundary conditions that the Hubble rates on the initial and final boundary hypersurfaces are fixed. Then, we modified the action to include a complex part such that when the TCC criteria are violated, the probability distribution of the universe to evolve into such states exponentially diminishes.\par
In evaluating the path integral, we found that the action we deal with has two real-valued saddle points. However, when quantized, perturbations against one of these saddle points lead to instability. Even though, mathematically, the Picard-Lefschetz theory allows for a contour of integration that picks up the contribution from both the saddle points in the semiclassical approximation leading to a real-valued cosinusoidal wave function much like the Hartle-Hawking's no-boundary wave function, the instability of perturbations around one of the saddle points forces us to restrict the range of lapse integration leading to a complex-valued wave function like the Vilenkin tunneling proposal.\par
As a consequence of the modification of the wave function with an exponential suppression factor that subdues the probability of the universe's evolution into a state that violates the TCC, we found that the universe is likely to have a characteristic Hubble rate at the end of inflation which is lower in value than what is classically expected. At the end of inflation, this characteristic Hubble rate introduces non-trivial scale non-invariant corrections to the power spectrum of quantum perturbations around the classical background geometry. However, in the limit of the absence of quantum gravity effects, we can still recover the expected scale-invariant power spectrum of the inflationary paradigm, showing the consistency of our results.\par
Though our analysis has been insightful, we still have worked in a de Sitter toy model of the universe without any matter content but a cosmological constant. In future investigations, our goal would be to include more realistic models of matter fields that drive the inflation and see what modifications to the wave function of the universe are expected to arise due to the Trans-Planckian Censorship Conjecture.

\acknowledgments

We gratefully acknowledge Sumanta Chakraborty and Karthik Rajeev for all the insightful discussions, which have led to a better understanding of the problem and greatly improved this work's content. This research is funded by the INSPIRE fellowship from the DST, Government of India (Reg. No. DST/INSPIRE/03/2019/001887).


%
\end{document}